\documentclass[final,5p,times]{elsarticle}
\usepackage{amssymb}
\usepackage{graphicx}

\begin{document}

\begin{frontmatter}

\title{Low-lying quasiparticle excitations in strongly-correlated superconductors: An ansatz from BCS quasiparticle excitations?}

\author{Chung-Pin Chou}
\address{Condensed Matter Physics and Materials Science
Department, Brookhaven National Laboratory, Upton, New York 11973,
USA}

\begin{abstract}
The question about the existence of Bogoliubov's quasiparticles in
the BCS wave functions underneath Gutzwiller's projection is of
importance to strongly correlated systems. We develop a method to
examine the two-particle excitations of Gutzwiller-projected BCS
wave functions by using the variational Monte Carlo approach. We
find that the exact Gutzwiller-projected quasiparticle (GQP)
dispersions are quantitatively reproduced by the
Gutzwiller-projected Bogoliubov quasiparticles (GBQP) except the
regions where $d$-wave Cooper pairing is strong. We believe GBQP
provides a reasonable description to the low-energy excitations in
strongly correlated superconducting systems because GBQP becomes
more stable than GQP near the antinodes. In addition, the intimate
connection between Gutzwiller's projection and $d$-wave Cooper
pairing may also imply that strong correlations play a significant
role in the nodal-antinodal dichotomy seen by photoemission
experiments in cuprates.
\end{abstract}


\end{frontmatter}

\section{Introduction}
\label{intro}

Two of the most intriguing puzzles in the study of high $T_{c}$
cuprates are the unexpected non-BCS behavior and the
non-quasiparticle nature in the superconducting states and the
normal states, respectively \cite{VishikNJP10}. To resolve those
puzzles, the relevant low energy physics based on the projection out
of the degrees of freedom at high energy must be embedded in a doped
Mott insulator \cite{LeeRMP06,PhillipsRMP10}. The
Gutzwiller-projected BCS wave function is the appropriate
description of the superconducting state in cuprates
\cite{AndersonSci87}, while strong correlations make theoretical
approaches extremely difficult. However, based on the framework of
the Gutzwiller-projected states, the issues related to the
finite-temperature physics of cuprates are still unclear
\cite{JainPNAS09,ShastryPRB10}. Therefore, the first step in
studying the excitations is to understand the structure of low-lying
quasiparticle excitations.

It has been experimentally observed that the low-lying excitations
of superconducting cuprates resemble BCS Bogoliubov's quasiparticles
(BQP) \cite{MatsuiPRL03}. Also, many theoretical studies on the
Gutzwiller-projected BQP (GBQP) excitations have been presented few
years ago
\cite{OhtaPRL94,YunokiPRL05,YunokiPRB05,NavePRB06,YunokiPRB06,KYYangPRB06,CPCPRB06}.
The ansatz used for the GBQP excited states is based on the success
of the Gutzwiller-projected BCS wave function. Owing to exact
diagonalization results indicating the well defined BCS-like BQP as
low-energy excitations of the $t-J$ model \cite{OhtaPRL94}, the GBQP
excited state given in Eq.(\ref{e:equ3}) is expected to be a simply
renormalized BQP excitation despite lack of analytical proof. Relied
on the careful fitting simulations \cite{KYYangPRB06}, we found they
are quantitatively satisfied with the renormalized BQP picture. Even
so, we still have no knowledge of the exact Gutzwiller-projected
quasiparticles (GQP). On the other hand, to explain some unusual
features seen by angle-resolved photoemission spectroscopy (ARPES)
in cuprates, the extension beyond the single-mode approach has also
been studied \cite{TanPRL08}. Therefore, from a theoretical point of
view, the difference between GBQP excitations and GQP excitations
should be clarified.

Let us briefly summarize the key messages involved in this article.
We begin by detailing the procedure that constructs the two-particle
excitations by using the usual GBQP picture and the GQP excitations
in the variational Monte Carlo (VMC) calculation. We demonstrate the
projected two-particle excitation is reasonable to be constructed by
applying BQP operators to strongly correlated superconducting ground
states. It is noticed that there is the discrepancy between GQP and
GBQP near antinodal regions. This discrepancy arising from a close
relation between Gutzwiller's projection and $d$-wave pairing may
provide a clue to the causes of the nodal-antinodal dichotomy
observed by ARPES measurements \cite{ZhouPRL04,GarfNP11}.

\section{Theory}
\label{theory}

Let us begin by
\begin{eqnarray}
\hat{H}=-\sum_{i,j,\sigma}t_{ij}\tilde{c}_{i\sigma}^{\dag}\tilde{c}_{j\sigma}+J\sum_{\langle
i,j\rangle}\left(\mathbf{S}_{i}\cdot\mathbf{S}_{j}-\frac{1}{4}n_{i}n_{j}\right),
\label{e:equ1}
\end{eqnarray}
where the hopping $t_{ij}=t$, $t'$, and $t''$ for sites i and j
being the nearest, second-nearest, and third-nearest neighbors,
respectively. Other notations are standard. We restrict the electron
creation operators $\tilde{c}_{i\sigma}^{\dag}$ to the subspace
without doubly-occupied sites. In the following, the bare parameters
in the Hamiltonian are set to be $(t',t'',J)/t=(-0.3,0.15,0.3)$. Two
holes are doped into the extended $t-t'-t''-J$ model in $16\times16$
lattice.

The well-known candidate for the ground state
\cite{AndersonSci87,EdeggerAIP07,OgataRPP08,CPChouPRB08} is the
$d$-wave resonating-valence-bond ($d$-RVB) wave function with
Jastrow factors of the form
\begin{equation}
|\Phi_{0}\rangle=\hat{P}_{N_{e}}\hat{P}_{J}\hat{P}_{G}\prod_{\mathbf{k}}\left(u_{\mathbf{k}}+v_{\mathbf{k}}c^{\dagger}_{\mathbf{k}
\uparrow}c^{\dagger}_{-\mathbf{k}
\downarrow}\right)|0\rangle,\label{e:equ2}
\end{equation}
where the coefficients $u_{k}$ and $v_{k}$ are the BCS coherence
factors. The trial wave function has three projections:
$\hat{P}_{N_{e}}$ to fix the number of electrons $N_{e}$, the
Gutzwiller projector
$\hat{P}_{G}(=\prod_{i}\left(1-n_{i\uparrow}n_{i\downarrow}\right))$
to enforce no-doubly-occupied sites, and charge-charge Jastrow
factors $\hat{P}_{J}$ to repel neighboring holes (see the details in
Ref.\cite{CPChouPRB08}). It is not shown that the conclusions in
this work would not be changed by $\hat{P}_{J}$, and hence we will
ignore the charge-charge Jastrow factors in the following.

Even so, it is still not easy to construct the single-particle
excited state of the extended $t-J$ Hamiltonian in the canonical
ensemble. Based on Eq.(\ref{e:equ2}), however, the simplest way is
to define a single-particle excitation under Gutzwiller's projection
as
\begin{equation}
|\Phi_{\mathbf{k}\sigma}\rangle=\hat{P}_{N_{e}\pm1}\hat{P}_{G}\gamma_{\mathbf{k}\sigma}^{\dag}|BCS\rangle,\label{e:equ3}
\end{equation}
where
$\gamma_{\mathbf{k}\sigma}^{\dag}(=u_{\mathbf{k}}c_{\mathbf{k}\sigma}^{\dag}-\sigma
v_{\mathbf{k}}c_{-\mathbf{k}\bar{\sigma}})$ is the creation of the
BQP and $\sigma$ spin index ($\bar{\sigma}=-\sigma$). In what
follows, to avoid the confusion due to mixing Hilbert space of
different particle numbers, we introduce the partial particle-hole
transformation to change the representation from (c) to (df)
\cite{YokoyamaJPSJ88,CPChouPRB12}:
\begin{eqnarray}
\begin{array}{ccc}
  (c) &   & (df) \\
  c_{i\uparrow} & \rightarrow & f_{i}^{\dagger} \\
  c_{i\downarrow} & \rightarrow & d_{i}
\end{array}\label{e:equ4}
\end{eqnarray}
Two different particles, $d$ and $f$, are thus introduced instead of
down- and up-spin electrons (see the details in
Ref.\cite{CPChouPRB12}). We start from the wave function without
Gutzwiller's projection. First, the BCS wave function can be
transformed into the representation (df),
\begin{eqnarray}
|BCS\rangle\rightarrow\prod_{\mathbf{k}}\left(u_{\mathbf{k}}f^{\dagger}_{-\mathbf{k}}-v_{\mathbf{k}}d^{\dagger}_{-\mathbf{k}}\right)|0\rangle_{(df)},\label{e:equ5}
\end{eqnarray}
where the subscripts indicate different representations. Then the
single-particle BCS excited state is similarly transform into
\begin{eqnarray}
\gamma_{\mathbf{k}\uparrow}^{\dag}|BCS\rangle\rightarrow\prod_{\mathbf{q}\neq\mathbf{k}}\left(u_{\mathbf{q}}f^{\dagger}_{-\mathbf{q}}-v_{\mathbf{q}}d^{\dagger}_{-\mathbf{q}}\right)|0\rangle_{(df)},\label{e:equ6}
\end{eqnarray}
\begin{eqnarray}
\gamma_{-\mathbf{k}\downarrow}^{\dag}|BCS\rangle\rightarrow
d^{\dagger}_{-\mathbf{k}}f^{\dagger}_{-\mathbf{k}}\prod_{\mathbf{q}\neq\mathbf{k}}\left(u_{\mathbf{q}}f^{\dagger}_{-\mathbf{q}}-v_{\mathbf{q}}d^{\dagger}_{-\mathbf{q}}\right)|0\rangle_{(df)}.\label{e:equ7}
\end{eqnarray}
In the representation (df), the total particle number of the
single-particle BCS excitation is no longer confused with $N_{e}+1$
or $N_{e}-1$ like Eq.(\ref{e:equ3}) in the representation (c), but
fixed to $N-1$ in Eq.(\ref{e:equ6}) and $N+1$ in Eq.(\ref{e:equ7}).

On the other hand, we adopt the similar route to write down the
two-particle BCS excitation
$\gamma_{\mathbf{k}\uparrow}^{\dag}\gamma_{-\mathbf{k}\downarrow}^{\dag}|BCS\rangle$
in the representation (df):
\begin{eqnarray}
-\left(u_{\mathbf{k}}d^{\dagger}_{-\mathbf{k}}+v_{\mathbf{k}}f^{\dagger}_{-\mathbf{k}}\right)\prod_{\mathbf{q}\neq\mathbf{k}}\left(u_{\mathbf{q}}f^{\dagger}_{-\mathbf{q}}-v_{\mathbf{q}}d^{\dagger}_{-\mathbf{q}}\right)|0\rangle_{(df)}.\label{e:equ8}
\end{eqnarray}
According to the above equation, if we define two states
$|\Psi_{\mathbf{k}}^{d}\rangle$ and $|\Psi_{\mathbf{k}}^{f}\rangle$
as
\begin{eqnarray}
|\Psi_{\mathbf{k}}^{d}\rangle&\equiv&-d^{\dagger}_{-\mathbf{k}}\prod_{\mathbf{q}\neq\mathbf{k}}\left(u_{\mathbf{q}}f^{\dagger}_{-\mathbf{q}}-v_{\mathbf{q}}d^{\dagger}_{-\mathbf{q}}\right)|0\rangle_{(df)},\nonumber\\
|\Psi_{\mathbf{k}}^{f}\rangle&\equiv&f^{\dagger}_{-\mathbf{k}}\prod_{\mathbf{q}\neq\mathbf{k}}\left(u_{\mathbf{q}}f^{\dagger}_{-\mathbf{q}}-v_{\mathbf{q}}d^{\dagger}_{-\mathbf{q}}\right)|0\rangle_{(df)},\label{e:equ9}
\end{eqnarray}
the BCS ground state and the two-particle BCS excitation will be
obviously given by
\begin{eqnarray}
|BCS\rangle&=&v_{\mathbf{k}}|\Psi_{\mathbf{k}}^{d}\rangle+u_{\mathbf{k}}|\Psi_{\mathbf{k}}^{f}\rangle,\nonumber\\
\gamma_{\mathbf{k}\uparrow}^{\dag}\gamma_{-\mathbf{k}\downarrow}^{\dag}|BCS\rangle&=&u_{\mathbf{k}}|\Psi_{\mathbf{k}}^{d}\rangle-v_{\mathbf{k}}|\Psi_{\mathbf{k}}^{f}\rangle,\label{e:equ10}
\end{eqnarray}
respectively. Owing to
$\langle\Psi_{\mathbf{k}}^{f}|\Psi_{\mathbf{k}}^{d}\rangle=0$,
Eqs.(\ref{e:equ10}) simply represent that both the BCS ground state
and the first BCS excited state are able to be expanded by the two
orthonormal states, $|\Psi_{\mathbf{k}}^{d}\rangle$ and
$|\Psi_{\mathbf{k}}^{f}\rangle$.

Applying the similar idea in the projected case, we can write down
the $d$-RVB ground state and the corresponding first GQP excited
state $|\Phi_{\mathbf{k}}^{+}\rangle$ (shown in Eq.(\ref{e:equ15}))
by using the following two states: $|\Phi_{\mathbf{k}}^{d}\rangle$
and $|\Phi_{\mathbf{k}}^{f}\rangle$ given by
\begin{eqnarray}
|\Phi_{\mathbf{k}}^{d}\rangle&=&-\hat{P}_{N_{e}}\hat{P}_{G}|\Psi_{\mathbf{k}}^{d}\rangle,\nonumber\\
|\Phi_{\mathbf{k}}^{f}\rangle&=&\hat{P}_{N_{e}}\hat{P}_{G}|\Psi_{\mathbf{k}}^{f}\rangle.\label{e:equ11}
\end{eqnarray}
As well, we can continue the single-particle excitation shown in
Eq.(\ref{e:equ3}) to create two-particle GBQP excited state:
\begin{equation}
|\Phi_{\mathbf{k}}^{GBQP}\rangle=\hat{P}_{N_{e}}\hat{P}_{G}\gamma_{\mathbf{k}\uparrow}^{\dag}\gamma_{-\mathbf{k}\downarrow}^{\dag}|BCS\rangle.\label{e:equ12}
\end{equation}
Some details in the VMC calculation should be noticed. To avoid the
divergence from the nodes in the trial wave functions, the boundary
condition we use is the anti-periodic boundary condition along both
x and y directions. In order to achieve a reasonable acceptance
ratio, the simulations consist of a combination of one-particle
moves and two-particle moves. Variational parameters in the $d$-RVB
state are optimized by using the stochastic reconfiguration method
\cite{SorellaPRB01}. All physical quantities are calculated with the
optimized parameters. We also take a sufficient number of samples
($=2\times10^{5}$) to reduce the statistical errors, and keep the
sampling interval ($\sim40$) long enough to ensure statistical
independence between samples.

\begin{figure}[top]
\begin{center}\rotatebox{0}{\includegraphics[height=3.7in,width=3.3in]{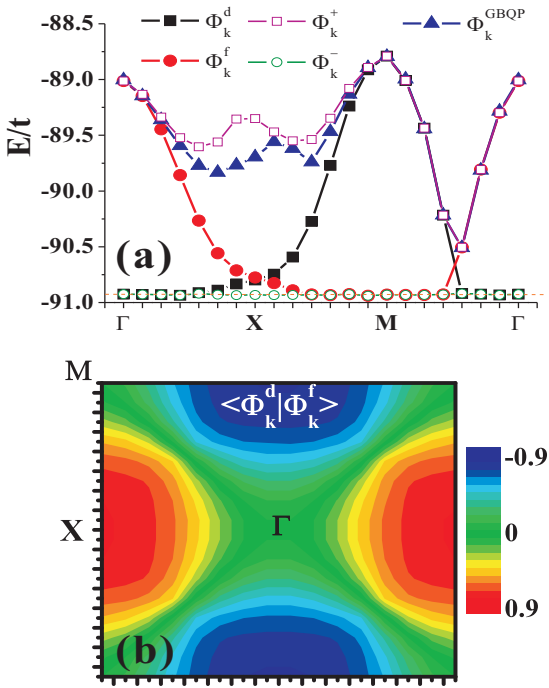}}\end{center}
\caption{(a) Two-particle dispersions of different excited states
for the $t-t'-t''-J$ model doping two holes in $16\times16$ lattice.
The orange dashed line denotes the $d$-RVB state
($E_{dRVB}=-90.93t$). The symbols for momenta:
$\Gamma\equiv(0.5,0.5)$, $X\equiv(5.5,0.5)$ and $M\equiv(5.5,5.5)$
in terms of $\pi/8$. (b) The momentum-space contour plot of the
overlap between $|\Phi_{k}^{d}\rangle$ and
$|\Phi_{k}^{f}\rangle$.}\label{fig1}
\end{figure}

\section{Results and discussion}
\label{result}

Since the projections are included in Eqs.(\ref{e:equ11}), the trial
states $|\Phi_{\mathbf{k}}^{d}\rangle$ and
$|\Phi_{\mathbf{k}}^{f}\rangle$ are no longer orthogonal. We need to
diagonalize a $2\times2$ Hamiltonian matrix in the subspace spanned
by Eqs.(\ref{e:equ11}). In principle, we can reconstruct two
orthogonal states, $|\hat{\Phi}_{\mathbf{k}}^{d}\rangle$ and
$|\hat{\Phi}_{\mathbf{k}}^{f}\rangle$, for each momentum by using
Gram-Schmidt method. The Hamiltonian matrix in this subspace is
given by
\begin{equation}
\hat{H}_{sub}=\left(
        \begin{array}{cc}
          H_{dd} & H_{df} \\
          H_{fd} & H_{ff} \\
        \end{array}
      \right),\label{e:equ13}
\end{equation}
where
$H_{ij}\equiv\langle\hat{\Phi}_{\mathbf{k}}^{i}|\hat{H}|\hat{\Phi}_{\mathbf{k}}^{j}\rangle$
and $i,j=d$ or $f$. We further diagonalize $\hat{H}_{sub}$ to obtain
the eigenstates $|\Phi_{\mathbf{k}}^{-}\rangle$ and
$|\Phi_{\mathbf{k}}^{+}\rangle$ as a linear combination of
$|\hat{\Phi}_{\mathbf{k}}^{d}\rangle$ and
$|\hat{\Phi}_{\mathbf{k}}^{f}\rangle$. Without the projection,
Eq.(\ref{e:equ10}) provides a route to construct the BCS ground
state and the two-particle excited state by using a linear
combination of two orthonormal states
$|\Psi_{\mathbf{k}}^{d}\rangle$ and $|\Psi_{\mathbf{k}}^{f}\rangle$.
Therefore, we can expect the ground state
$|\Phi_{\mathbf{k}}^{-}\rangle$ and the GQP excited state
$|\Phi_{\mathbf{k}}^{+}\rangle$ should play the same role in the
cases under the projection. A further question is whether the GBQP
wave function $|\Phi_{\mathbf{k}}^{GBQP}\rangle$ mentioned above is
appropriate to describe the two-particle excitation from the $d$-RVB
state $|\Phi_{0}\rangle$. In Fig.\ref{fig1}(a), we compare the
dispersion of four different states discussed above with
$|\Phi_{\mathbf{k}}^{GBQP}\rangle$. First, it is obvious that
$|\Phi_{\mathbf{k}}^{-}\rangle$ exactly reproduces the optimized
energy of the $d$-RVB state indicated by the orange dashed line.
Based on this agreement, we are confident that
$|\Phi_{\mathbf{k}}^{+}\rangle$ should properly represent the
two-particle GQP excitation. Interestingly, except the deviation
near the antinodal regions, the energy dispersion of
$|\Phi_{\mathbf{k}}^{+}\rangle$ coincides with
$|\Phi_{\mathbf{k}}^{GBQP}\rangle$ very well. We shall return to
this deviation in the following. Here we see that
$|\Phi_{\mathbf{k}}^{GBQP}\rangle$ has the lower energy than
$|\Phi_{\mathbf{k}}^{+}\rangle$ around the regions with large
$d$-wave pairing amplitude. Therefore, we may conclude that the GQP
excitation seems reasonable to be constructed by applying BQP
operators to the $d$-RVB wave function although
$|\Phi_{\mathbf{k}}^{+}\rangle$ just represents the first
two-particle GQP excited state.

Second, we notice that $|\Phi_{\mathbf{k}}^{d}\rangle$
($|\Phi_{\mathbf{k}}^{f}\rangle$) shows dispersionless behavior
inside (outside) the underlying Fermi surface. This results can be
easily understood by transforming the representation (df) back to
(c). In the original representation (c), they are given by
\begin{eqnarray}
|\Phi_{\mathbf{k}}^{d}\rangle&=&\hat{P}_{N_{e}}\hat{P}_{G}c^{\dagger}_{\mathbf{k}
\uparrow}c^{\dagger}_{-\mathbf{k}\downarrow}\prod_{\mathbf{q\neq
k}}\left(u_{\mathbf{q}}+v_{\mathbf{q}}c^{\dagger}_{\mathbf{q}\uparrow}c^{\dagger}_{-\mathbf{q}\downarrow}\right)|0\rangle_{(c)},\nonumber\\
|\Phi_{\mathbf{k}}^{f}\rangle&=&\hat{P}_{N_{e}}\hat{P}_{G}\prod_{\mathbf{q\neq
k}}\left(u_{\mathbf{q}}+v_{\mathbf{q}}c^{\dagger}_{\mathbf{q}\uparrow}c^{\dagger}_{-\mathbf{q}\downarrow}\right)|0\rangle_{(c)}.\label{e:equ14}
\end{eqnarray}
Apparently, there is an additional electron (hole) pair in
$|\Phi_{\mathbf{k}}^{d}\rangle$ ($|\Phi_{\mathbf{k}}^{f}\rangle$) so
that it only can show the dispersion outside (inside) the underlying
Fermi surface in Fig.\ref{fig1}(a). Next, we study the overlap
between $|\Phi_{\mathbf{k}}^{d}\rangle$ and
$|\Phi_{\mathbf{k}}^{f}\rangle$ shown in Fig.\ref{fig1}(b). Note
that without the projection, $|\Psi_{\mathbf{k}}^{d}\rangle$ and
$|\Psi_{\mathbf{k}}^{f}\rangle$ are orthogonal for every momentum.
Under the projection the overlap dramatically enhances near the
antinodal regions. The shape of the overlap in the momentum space is
very similar to the $d$-wave form factor,
$\cos\mathbf{k_{x}}-\cos\mathbf{k_{y}}$, and the maxima about $0.82$
right at antinodes. It can be easily understood from
Eq.(\ref{e:equ14}) in the original representation (c). Since there
is an extra Cooper pair $c^{\dagger}_{\mathbf{k}
\uparrow}c^{\dagger}_{-\mathbf{k}\downarrow}$ in the overlap, we can
expect that Gutzwiller's projection would influence the Cooper pair
following the $d$-wave symmetry in BCS wave functions. Thus, we can
further comprehend the deviation for the excitation energy near the
antinodal regions in Fig.\ref{fig1}(a).

\begin{figure}[top]
\begin{center}\rotatebox{0}{\includegraphics[height=3.5in,width=2.5in]{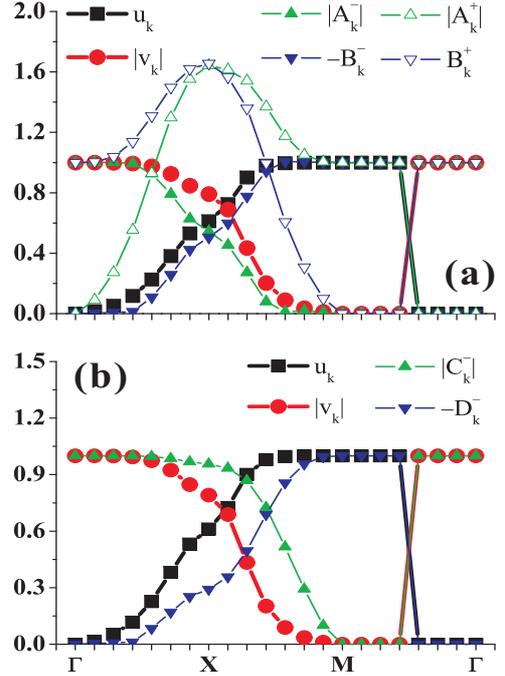}}\end{center}
\caption{The coefficients (a) $A_{\mathbf{k}}^{\pm}$ and
$B_{\mathbf{k}}^{\pm}$; (b) $C_{\mathbf{k}}^{\pm}$ and
$D_{\mathbf{k}}^{\pm}$ along the high-symmetric momenta. Their
definitions are explained in the text. The BCS coherence factors
$u_{k}$ and $v_{k}$ obtained from the optimized parameters are shown
for comparison.}\label{fig2}
\end{figure}

Furthermore, it is important to examine the details of the
eigenstates $|\Phi_{\mathbf{k}}^{\pm}\rangle$. We can write down
$|\Phi_{\mathbf{k}}^{\pm}\rangle$ as a linear combination of the
orthogonal states $|\hat{\Phi}_{\mathbf{k}}^{d/f}\rangle$:
\begin{eqnarray}
|\Phi_{\mathbf{k}}^{\pm}\rangle=C_{\mathbf{k}}^{\pm}|\hat{\Phi}_{\mathbf{k}}^{d}\rangle+D_{\mathbf{k}}^{\pm}|\hat{\Phi}_{\mathbf{k}}^{f}\rangle,\label{e:equ15}
\end{eqnarray}
where the coefficients $C_{\mathbf{k}}^{\pm}$ and
$D_{\mathbf{k}}^{\pm}$ can be easily determined by diagonalizing
Eq.(\ref{e:equ13}). By using Gram-Schmidt method, the two orthogonal
states $|\hat{\Phi}_{\mathbf{k}}^{d/f}\rangle$ can be written as
\begin{eqnarray}
|\hat{\Phi}_{\mathbf{k}}^{d}\rangle&=&|\Phi_{\mathbf{k}}^{d}\rangle,\nonumber\\
|\hat{\Phi}_{\mathbf{k}}^{f}\rangle&=&\frac{|\Phi_{\mathbf{k}}^{f}\rangle-\langle\Phi_{\mathbf{k}}^{d}|\Phi_{\mathbf{k}}^{f}\rangle|\Phi_{\mathbf{k}}^{d}\rangle}{\sqrt{1-|\langle\Phi_{\mathbf{k}}^{d}|\Phi_{\mathbf{k}}^{f}\rangle|^{2}}}.\label{e:equ16}
\end{eqnarray}
Thus, the eigenstates $|\Phi_{\mathbf{k}}^{\pm}\rangle$ are
expressed in terms of $|\Phi_{\mathbf{k}}^{d/f}\rangle$ as well,
\begin{eqnarray}
|\Phi_{\mathbf{k}}^{\pm}\rangle=A_{\mathbf{k}}^{\pm}|\Phi_{\mathbf{k}}^{d}\rangle+B_{\mathbf{k}}^{\pm}|\Phi_{\mathbf{k}}^{f}\rangle.\label{e:equ17}
\end{eqnarray}
Here the coefficients $A_{\mathbf{k}}^{\pm}$ and
$B_{\mathbf{k}}^{\pm}$ are related to $C_{\mathbf{k}}^{\pm}$ and
$D_{\mathbf{k}}^{\pm}$ according to Eq.(\ref{e:equ16}). Note that
the normalization condition guarantees
$|C_{\mathbf{k}}^{\pm}|^{2}+|D_{\mathbf{k}}^{\pm}|^{2}=1$ but not
necessary for $A_{\mathbf{k}}^{\pm}$ and $B_{\mathbf{k}}^{\pm}$.

In order to understand how Gutzwiller's projection affects the
coherence factors in Eq.(\ref{e:equ10}), we plot the comparison
between the BCS coherence factors and the coefficients
$A_{\mathbf{k}}^{\pm}$ and $B_{\mathbf{k}}^{\pm}$ in
Fig.\ref{fig2}(a). We should notice that $u_{\mathbf{k}}$ and
$v_{\mathbf{k}}$ are obtained from the optimized variational
parameters. To avoid the confusion arising from the sign of $d$-wave
symmetry, we show the absolute value of $v_{\mathbf{k}}$ and
$A_{\mathbf{k}}^{\pm}$. Along the nodal direction, interestingly,
the projection seems not to change the BCS coherence factors. This
may imply the intimate relation between Gutzwiller's projection and
the $d$-wave gap function. However, since the coefficients
$A_{\mathbf{k}}^{\pm}$ and $B_{\mathbf{k}}^{\pm}$ are not
normalized, there is a large enhancement for the first GQP excited
state and small suppression for the $d$-RVB ground state around the
antinodal parts. In Fig.\ref{fig2}(b), we clearly demonstrate the
shape of the coefficients $C_{\mathbf{k}}^{\pm}$ and
$D_{\mathbf{k}}^{\pm}$ looks similar to the BCS coherence factors in
spite of the existence of Gutzwiller's projection. The crossing
curve of $C_{\mathbf{k}}^{\pm}$ and $D_{\mathbf{k}}^{\pm}$ bends
more like a hole pocket at $(\pi,\pi)$, and however the BCS
coherence factor displays a diamond-like underlying Fermi surface
(not shown). Even so, it is reasonable to believe that the GQPs in
the $d$-RVB wave function are still analogous to the BQP picture.

\section{Conclusion}
\label{conclusion}

In conclusions, we have developed a method to examine the idea of
BQPs in Gutzwiller-projected BCS wave functions. By calculating the
two-particle excitation dispersion using VMC approach, the GQP
excited states have been obtained. We have found the GBQP
$|\Phi_{\mathbf{k}}^{GBQP}\rangle$ shows almost the same energy as
the GQP $|\Phi_{\mathbf{k}}^{+}\rangle$ except that gives the lower
energy around the antinodal regime. The reason is that there exists
large overlap between $|\Phi_{\mathbf{k}}^{d}\rangle$ and
$|\Phi_{\mathbf{k}}^{f}\rangle$ near the antinodes, suggesting the
intimate connection between Gutzwiller's projection and $d$-wave
Cooper pairing. It also results in the deviation of the coefficients
in $|\Phi_{\mathbf{k}}^{\pm}\rangle$ from the BCS coherence factors,
which might be related to the nodal-antinodal dichotomy observed in
cuprates by ARPES measurements. Therefore, the lower energy around
the antinodal regions implies the GBQP excited wave functions are
suitable to describe the low-lying excitations in strongly
correlated superconductors.

\section{Acknowledgments}
\label{Acknowledgment}

The author thanks T.-K. Lee and Wei Ku for useful discussions. This
work is supported by Brookhaven Science Associates, LLC under
Contract No. DE-AC02-98CH10886 with the U.S. Department of Energy
and the Postdoctoral Research Abroad Program sponsored by National
Science Council in Taiwan with Grant No. NSC 101-2917-I-564-010.
\\

\end{document}